\documentclass[a4paper, 10pt, conference]{IEEEtran}
\IEEEoverridecommandlockouts                   

\usepackage{graphicx}
\usepackage{color}
\usepackage{amsmath}
\usepackage{multirow}
\usepackage{booktabs}
\usepackage{url}
\usepackage{amsmath}
\usepackage[position=t,singlelinecheck=off]{subfig}
\usepackage{adjustbox}
\usepackage{subfloat}
\usepackage[english]{babel}
\usepackage{comment}
\usepackage{tikz}
\usetikzlibrary{positioning}

\newcommand\copyrighttext{%
  \footnotesize \textcopyright 2020 IEEE. Personal use of this material is permitted. Permission from IEEE must be obtained for all other uses, in any current or future media, including reprinting/republishing this material for advertising or promotional purposes, creating new collective works, for resale or redistribution to servers or lists, or reuse of any copyrighted component of this work in other works. This paper has been accepted for publication in the 2020 Twelfth International Conference on Quality of Multimedia Experience (QoMEX).}
\newcommand\copyrightnotice{%
\begin{tikzpicture}[remember picture,overlay]
\node[anchor=south,yshift=10pt] at (current page.south) {\fbox{\parbox{\dimexpr\textwidth-\fboxsep-\fboxrule\relax}{\copyrighttext}}};
\end{tikzpicture}%
}

\begin{document}

\title{Influence of Hand Tracking as a way of Interaction in Virtual Reality on User Experience}

\author{
 \IEEEauthorblockN{Jan-Niklas Voigt-Antons$^{1,2}$, Tanja Kojic$^1$, Danish Ali$^1$, Sebastian M\"oller$^{1,2}$}
 \IEEEauthorblockA{$^1$Quality and Usability Lab, TU Berlin, Germany\\
 $^2$German Research Center for Artificial Intelligence (DFKI), Berlin, Germany}
}


\maketitle
\copyrightnotice

\begin{abstract}
With the rising interest in Virtual Reality and the fast development and improvement of available devices, new features of interactions are becoming available. One of them that is becoming very popular is hand tracking, as the idea to replace controllers for interactions in virtual worlds.
This experiment aims to compare different interaction types in VR using either controllers or hand tracking. Participants had to play two simple VR games with various types of tasks in those games - grabbing objects or typing numbers. While playing, they were using interactions with different visualizations of hands and controllers. The focus of this study was to investigate user experience of varying interactions (controller vs. hand tracking) for those two simple tasks.
Results show that different interaction types statistically significantly influence reported emotions with Self-Assessment Manikin (SAM), where for hand tracking participants were feeling higher valence, but lower arousal and dominance. Additionally, task type of grabbing was reported to be more realistic, and participants experienced a higher presence.
Surprisingly, participants rated the interaction type with controllers where both where hands and controllers were visualized as statistically most preferred.
Finally, hand tracking for both tasks was rated with the System Usability Scale (SUS) scale, and hand tracking for the task typing was rated as statistically significantly more usable.
These results can drive further research and, in the long term, contribute to help selecting the most matching interaction modality for a task.
\end{abstract}



\section{INTRODUCTION \& RELATED WORK}
The recent popularity of Virtual Reality (VR) is resulting not only in many new applications available on the market but also in many new improvements and new features of head-mounted displays (HMD). One of the features of virtual experiences is to immerse users into other worlds \cite{zahorik1998presence}. Any contact with the real world such as sensors or controllers can break this feeling \cite{mcgloin2013video}. Therefore, new HMDs are allowing features of interacting via hand tracking to replace interactions with controllers.\footnote{https://www.oculus.com/blog/thumbs-up-hand-tracking-now-available-on-oculus-quest} This new possibility can lead to very interesting use cases in virtual reality in the field of gaming, but also in others fields where virtual reality is used for serious games. Virtual reality is already being widely used in learning \cite{monahan2008virtual}, more particularly in medicine \cite{mcgrath2018using} or manufacturing \cite{nee2013virtual}. Those fields are traditionally associated with precision, as well as hand movements are important. However, it is also important to ensure good quality and user experience to achieve good simulations and results.  
Based on the recent release of the build-in hand tracking technology by the company Oculus for their HMD Quest and the fact that there is still little to no scientific studies performed to test for influences on user experience, this paper aims to investigate following questions: 1) How do different interaction types (such as hand tracking or interaction via controllers) in virtual reality influence user experience?, and 2) Do different task types (such as grabbing or typing) for those interaction types have an influence on user experience? The feeling of presence and immersion are critical factors when it comes to simulations in VR, and a lot of research has been done already about it \cite{mania2001effects}, \cite{brooks1999s}. An essential element in the user experience of virtual reality is to accurately represent the user's hand in the virtual environment \cite{manresa2005hand}, \cite{gratch2002creating}.
This not only related to virtual reality but has already been researched in the field of gaming with other types of virtual environments \cite{wang2009real}, putting in focus in particular algorithms on how to achieve precise hand tracking via various sensors \cite{penelle2014multi}, \cite{taylor2016efficient}.
There is a study reporting on hand tracking and visualization in a virtual reality simulation where a design had been developed offering to track the fingers and palm \cite{cameron2011hand}. The study was focusing on experiments where participants had to mimic particular hand positions and reports on performance in terms of completion time. 
However, little work has been done with the focus on user experience with hand tracking in virtual reality. To the best of our knowledge, there is none research done with build-in hand tracking in HMD focusing on user experience with different interaction types.  

\section{METHODS}
\begin{figure}
    \centering
    \includegraphics[width=0.4\textwidth]{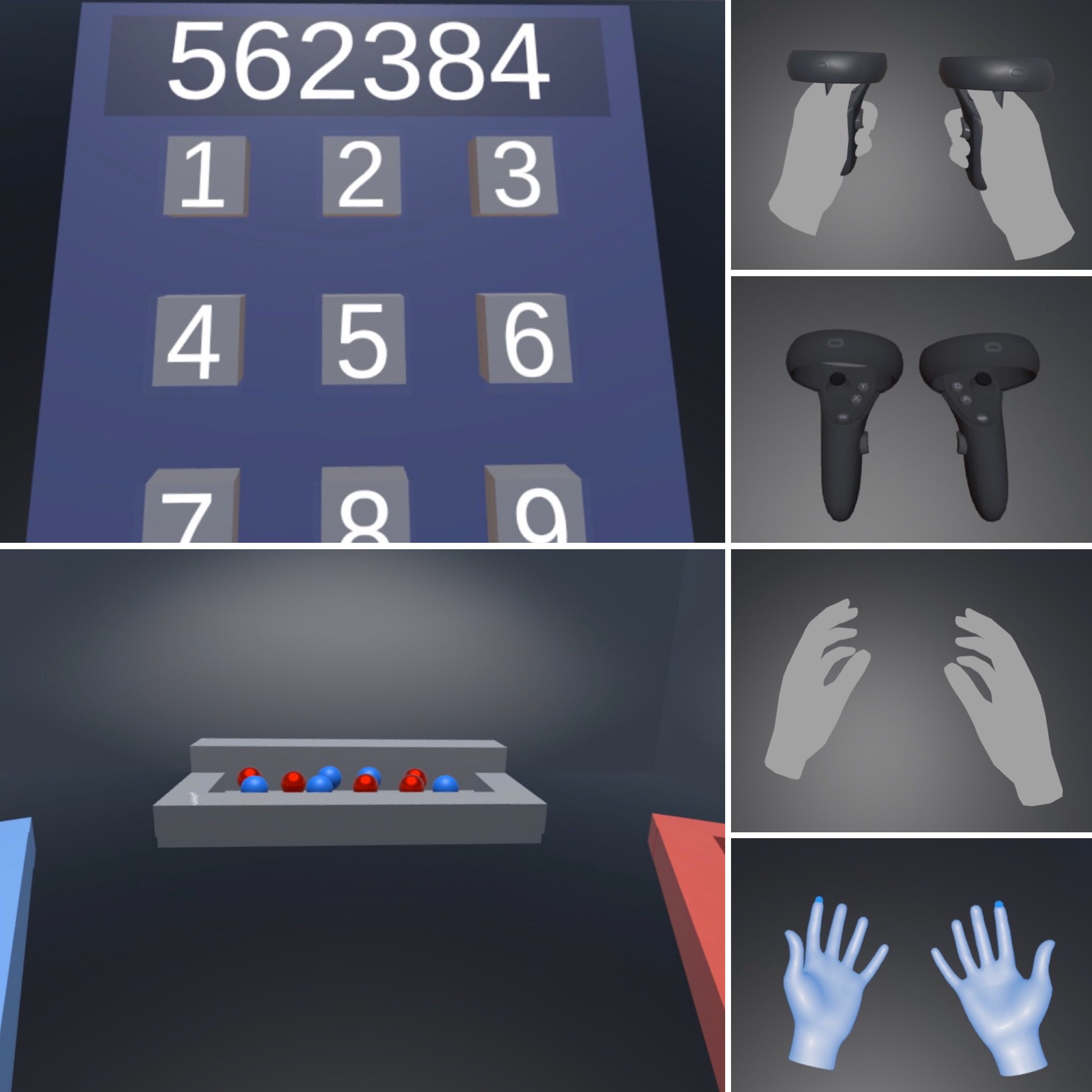}
    \caption{Left: Preview of typing task (upper) and preview of grabbing task (below); Right (top to bottom): interaction with visualization of both controllers and hands, visualization of only controllers, visualization of only hands, and interaction with hand tracking.}
    \label{fig:apps}
    \vspace{-2em}
\end{figure}

The study was conducted in a separate university's lab room equipped with Oculus Quest head-mounted display (HMD). This device was used in this study as it is supporting interactions using hand tracking as well as interactions with controllers. For this study, two Unity game applications had been created to test two different types of tasks - grabbing and typing. 
The first application was a game with a goal to sort five red and five blue balls by grabbing them and putting them to the correct box, as shown in Figure \ref{fig:apps}. The second application was a game where the player had to retype displayed number sequences on a virtual numerical keyboard, as shown in Figure \ref{fig:apps}, and repeat it for five courses. 
Both games had a strong focus to be played with hands, depending on the type of the game to grab or to type, and both games had the same four possible types of interactions: 1) Visualized controllers and hands, 2) Visualized controllers, but no visualization of hands, 3) Visualized hands, but no visualization of controllers, 4) Visualized hands while using hand tracking.
All visualizations, for both interaction types with controllers and with hand tracking, are shown in Figure \ref{fig:apps}. 

At the beginning of the experiment, participants were welcomed. As part of the introduction to participants, it was explained that the assignment is to play two VR games with different tasks, using controllers and hand tracking as types of interaction. Participants also signed a consent form, and before any VR condition, they were given a questionnaire asking about demographics, their previous experience with virtual reality, and hand tracking. Also, participants were asked about their tendency to actively engage in intensive technology interaction by Affinity for Technology Interaction (ATI) Scale questionnaire \cite{franke2019personal}. 
As part of the introduction to virtual reality and interactions, participants were explained how to use controllers and hand tracking for interaction, and they were given a chance to familiarize themselves with virtual environments of both games. After they confirmed to the moderator that they feel comfortable with learned interactions, the moderator could start the experiment. Each participant was playing two games with different task types grabbing or typing four times with each of the interaction types, which means that each participant overall participated in 8 conditions. The moderator asked the participants to observe the visualization and behavior of the interaction at the beginning of each condition. After each condition, participants were asked to fill in the Self-Assessment Manikin (SAM) \cite{bradley1994measuring} as an emotion assessment tool that uses graphic scales. Additionally, after each condition, participants were also rating feeling of presence and realism using a standardized measurement of perceived presence with a subjective scale: the Igroup Presence Questionnaire (IPQ) \cite{Schubert2001}. After all conditions, the participants were asked to rate the usability of reading in virtual reality with the System Usability Scale (SUS) \cite{brooke1996sus} for hand tracking separately for both grabbing and typing task. Finally, at the end, participants were asked to fill in a final questionnaire asking to rank the four different interactions in their experience from best to worse.
In total, there were 20 participants, eight female, and 12 male, with the average age of the participants was 30.20 years (SD = 12.74, min = 19, max = 62). When it comes to previous experience with VR technologies, 4 participants had no experience, while the others tried VR before, but nobody reported to be an expert. Further on, 9 participants reported having no experience at all with hand tracking for interactions, while others had little or some experience, and nobody was an expert with hand tracking. 
Data collection was conducted following ethical principles for medical research involving human subjects.

\section{RESULTS}
A repeated measure Analysis of Variance (ANOVA) was performed to determine statistically significant differences to investigate the effects between conditions with different task types and interaction types. The two tasks in the application were grabbing and typing, which had to be fulfilled using different interaction types (using controllers with visualization of only controllers, controllers with visualization of only hands, controllers with visualization of hands and controllers or using hand tracking). Further on, the Friedman's Two-Way Analysis was done to find out significant differences in preference of interaction types. Table \ref{tab:results} and \ref{tab:results2} provide an overview of the statistically significant results of the Friedman and the repeated measure ANOVA tests.

\begin{table}[htbp]
  \centering
  \caption{Effects of both task types (grabbing and typing), and all interaction types (hand tracking or controllers with different visualizations of hands and controllers) on arousal (SAM\_A), valence (SAM\_V) and dominance (SAM\_D) measured with SAM, and feeling of presence (PRES) and realism (REAL) measured with IPQ.}
  \resizebox{\columnwidth}{!}{
    \begin{tabular}{llcllrl}
    \toprule
\multicolumn{1}{l}{Parameter} & \multicolumn{1}{l}{Effect} & \multicolumn{1}{c}{$df_{\textnormal n}$} & \multicolumn{1}{l}{$df_{\textnormal d}$} & \multicolumn{1}{c}{$F$} & \multicolumn{1}{c}{$p$} & \multicolumn{1}{c}{$\eta_{\textnormal G}^2$}
    \\
    \midrule
   Task & SAM\_A & $1$     & $20$    & $10.35$ & $ .005$ & $0.35$ \\
   Task & PRES & $1$     & $20$    & $5.30$ & $ .033$ & $0.22$ \\
   Task & REAL & $1$     & $20$    & $4.90$ & $ .039$ & $0.21$ \\
   Interaction & SAM\_A & $3$     & $20$    & $5.25$ & $ .003$ & $0.22$ \\
   Interaction & SAM\_V & $3$     & $20$    & $20.08$ & $ <.001$ & $0.51$ \\
   Interaction & SAM\_D & $3$     & $20$    & $11.43$ & $ <.001$ & $0.38$ \\
    \bottomrule
    \end{tabular}
    }
  \label{tab:results}
  \vspace{-1.5em}
\end{table}

\begin{table}[htbp]
  \centering
  \tiny
  \caption{Post hoc analysis with Wilcoxon signed-rank tests for significantly different effects of ranking for all interaction types: Controllers with visualization of both hands and controllers (C\_H), controllers with only visualizations of hands (C\_OnlyH), controllers with visualizations of only controllers (C\_noH) and hand tracking (HTrack).}
  \resizebox{\columnwidth}{!}{
    \begin{tabular}{llcllrl}
    \toprule
 \multicolumn{1}{l}{Interaction} & \multicolumn{1}{c}{$df_{\textnormal n}$} & \multicolumn{1}{l}{$df_{\textnormal d}$} & \multicolumn{1}{c}{ $Z$ } & \multicolumn{1}{c}{$p$} & 
    \\
    \midrule
    C\_H and C\_OnlyH & $3$     & $20$    & $-3.079$ & $ .002$  \\
    C\_H and C\_noH & $3$     & $20$    & $-3.162$ & $ .002$  \\
    C\_H and HTrack & $3$     & $20$    & $-2.844$ & $ .004$  \\

    \bottomrule
    \end{tabular}
    }
  \label{tab:results2}
  \vspace{-1.5em}
\end{table}

\begin{figure}
    \centering
    \includegraphics[width=0.5\textwidth]{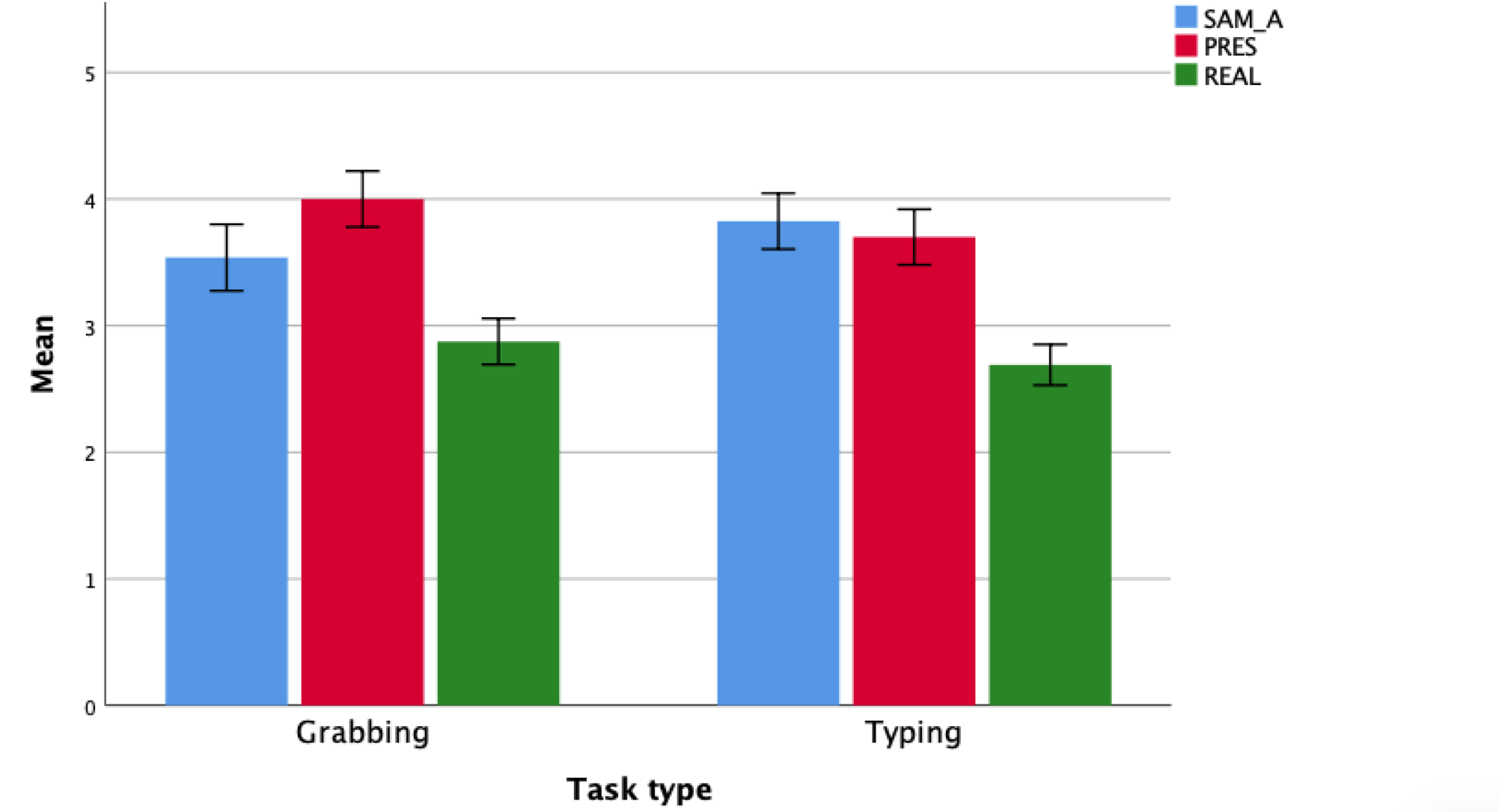}
    \caption{Mean values of SAM arousal dimension (SAM\_A), presence (PRES), and realism (REAL) over all participants for both tasks grabbing and typing. Whiskers denote 95 percent confidence intervals.}
    \label{fig:task}
    \vspace{-2em}
\end{figure}

\begin{figure}
    \centering
    \includegraphics[width=0.5\textwidth]{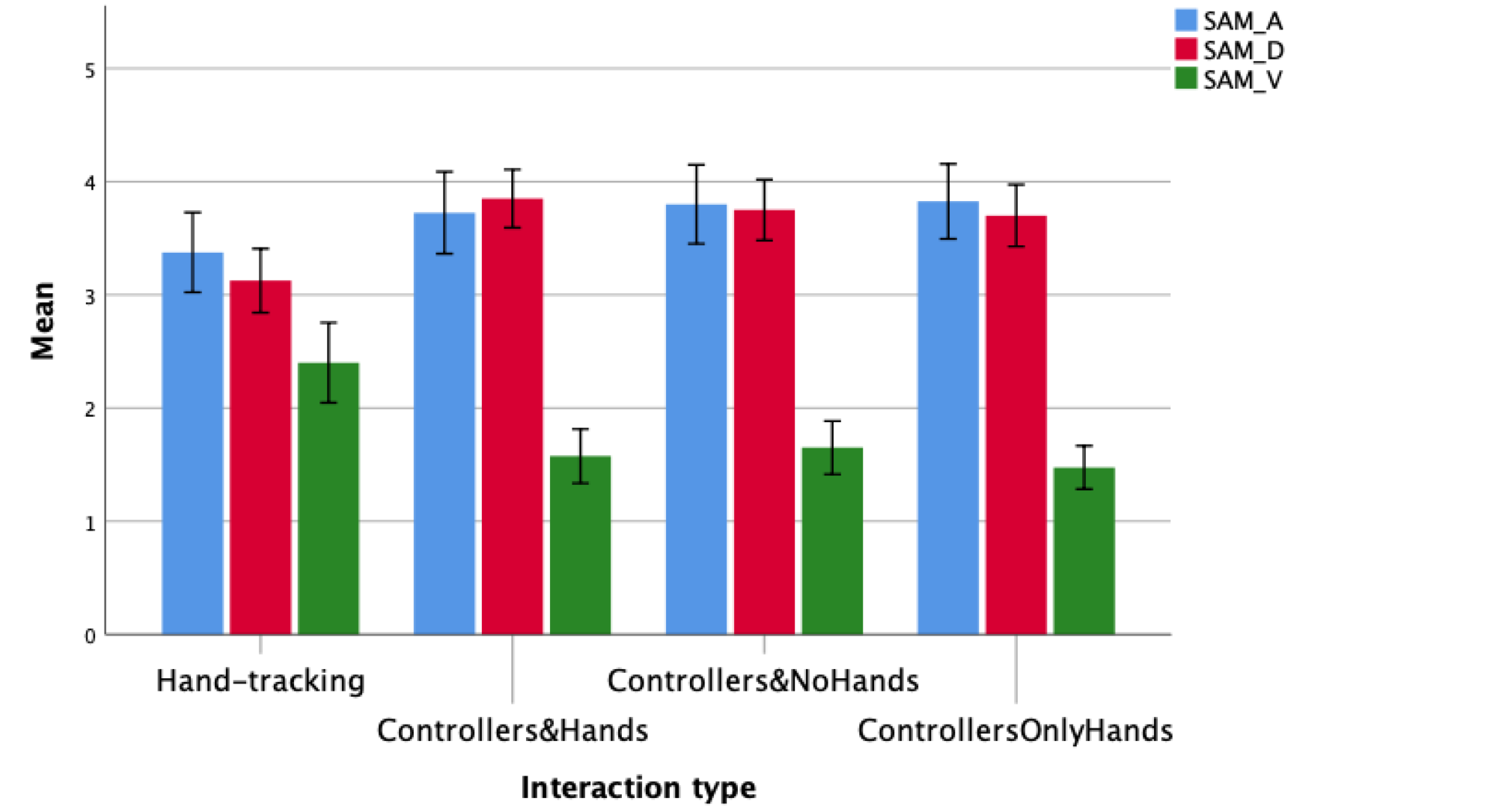}
    \caption{Mean values of the SAM dimensions: arousal (SAM\_A),valence (SAM\_V), and dominance (SAM\_D) over all participants for all interaction types with controllers (visualised as both controllers and hands, only controller, only hands) and hand tracking. Whiskers denote 95 percent confidence intervals.}
    \label{fig:inter}
    \vspace{-2em}
\end{figure}

There were statistically significant differences due to the conditions on perceived arousal, presence, and realism (see Table \ref{tab:results} and Figure \ref{fig:task}). 
For the different conditions where participants had different task types, either grabbing or typing, it can be observed that it influenced how the users have rated SAM dimension arousal. For the typing task, arousal was rater as significantly higher (M\,=\,3.83, SE\,=\,0.21) compared to the grabbing task (M\,=\,3.54, SE\,=\,0.22). 
Further on, participants have reported feeling significantly higher sense of presence while playing the task where they had to grab the balls (M\,=\,4.00, SE\,=\,0.15) compared to the task where they were typing (M\,=\,3.70, SE\,=\,0.17). Similarly, the feeling of realism was significantly higher for grabbing task (M\,=\,2.88, SE\,=\,0.14) compared to the typing task (M\,=\,2.69, SE\,=\,0.14). 
Comparing influences that were caused by different interaction types (with controllers or hand tracking), for all three dimensions of SAM questionnaire, statistically significant differences were found (see Table \ref{tab:results} and Figure \ref{fig:inter}). 
SAM dimension of arousal was rated significantly higher for interaction type using only hand tracking (M\,=\,3.38, SE\,=\,0.21) compared to the interaction with the controller where only hands were visualized (M\,=\,3.83, SE\,=\,0.21). 
Similarly, the dimension of valence was for hand tracking (M\,=\,2.40, SE\,=\,0.20) significantly higher compared to all other interaction types with controllers visualizes either as only controllers (M\,=\,1.65, SE\,=\,0.14), only hands (M\,=\,1.48, SE\,=\,0.12), or with visualization of both (M\,=\,1.58, SE\,=\,0.16). 
Interestingly, the dimension of dominance was reported to be significantly lower for hand tracking (M\,=\,3.13, SE\,=\,0.16) compared to all other conditions with visualization of both controllers and hands (M\,=\,3.85, SE\,=\,0.14), visualization of only controllers (M\,=\,3.75, SE\,=\,0.16), or visualization of only hands (M\,=\,3.70, SE\,=\,0.14). 

Ranking of the preferred way of interaction types, where participants had to order from best (4) interaction type to worst (1), resulted in statistically significant differences (see Table \ref{tab:results2}). 
A side corrected pairwise comparison showed that interaction type with controllers where both where hands and controllers were visualized (M\,=\,3.30, SE\,=\,0.47) were assigned a significantly higher rank compared to all other conditions: controllers with visualizations of only controllers (M\,=\,2.35, SE\,=\,0.93), of only hands (M\,=\,2.25, SE\,=\,1.16), and using hand tracking (M\,=\,2.10, SE\,=\,1.37).
A paired-samples t-test was conducted to compare average SUS scale values in hand tracking for grabbing (M\,=\,56.50, SE\,=\,18.09), and hand tracking for typing (M\,=\,75.62, SE\,=\,20.96). There was a significant difference in the scores; t(19)\,=\,3.01, p\,=\,0.007.

\section{DISCUSSION \& CONCLUSION}
The type of task (grabbing vs. typing) had an impact on how participants experienced the experimental condition. During the grabbing task, participants experienced lower arousal, higher presence, and higher realism. This result can be assigned to the fact that a grabbing task is more natural, therefore resulting in a higher realism and presence experience. The higher arousal of participants during the typing task could be due to higher demand during the more unnatural task.
The type of interaction (controllers visualized as both controllers and hands, the controller only, the controller only hands, and hand tracking) did not influence the experience of presence and realism. Nevertheless, there was an influence of interaction type on all SAM dimensions (valence, arousal, and dominance). It can be seen that participants felt less aroused, have a more positive experience (higher valence), and felt less dominant during the experimental condition hand tracking compared to all interaction types involving controllers. The higher valence for hand tracking can be addressed to the fact that users like the experience better. The lower arousal values could also be due to reduced demand due to the interactions imposed on the participants. The lower dominance experience could be to the fact that interacting with hand tracking leads to a lower feeling of control/precision and, therefore, also to a lower dominance rating. Finally, participants ranked the interaction type, including controllers visualized as both controllers and hands, as being more preferred than all other interaction types. The latter effect could be a combination of being used to interact with controllers and that most tasks are currently rather developed for typing selecting with a controller. Users still seem, when it comes to a visualization, preferred that not only the controller but also the hands are visualized. Finally, hand tracking for typing was rated statistically significantly more usable as hand tracking for the task grabbing on the SUS scale. This effect could be due to the fact that users are still more used to controller interaction and, therefore, also rate this experience a being more usable. This study showed that user experience using pure hand tracking interaction lead to a more positive and less arousing experience while reducing the sense of dominance/control. These results can drive further research and, in the long term, contribute to help selecting the most matching interaction modality for a task.

\bibliographystyle{IEEEtran}
\bibliography{references.bib} 

\end{document}